\documentclass[11pt]{article}
\usepackage{graphicx,smartdiagram}
\usepackage{amssymb}
\usepackage{amsmath}
\usepackage{amsthm}
\usepackage[symbol]{footmisc}
\usepackage{xcolor}

\def\be{\begin{equation}}
\def\ee{\end{equation}}
\newcommand{\ff}[1]{{{\boldsymbol{#1}}}} 
\def\x{{\ff{x}}}

\begin{document}

\title{Parameter Tuning of the Firefly Algorithm by Three Tuning Methods: Standard Monte Carlo, Quasi-Monte Carlo and Latin Hypercube Sampling Methods}

\author{Geethu Joy$^{1,2}$\thanks{Corresponding author},
Christian Huyck$^{1}$, Xin-She Yang$^{1}$}

\date{
1. School of Science and Technology, Middlesex University,
The Burroughs, London, NW4 4BT, UK \\[10pt]
2. Computer Engineering and Informatics, Middlesex University Dubai,
Dubai Knowledge Park, P. O. Box 500697, Dubai, United Arab Emirates
}

\maketitle

\begin{abstract}
There are many different nature-inspired algorithms in the literature, and almost all such algorithms have algorithm-dependent parameters that need to be tuned. The proper setting and parameter tuning should be carried out to maximize the performance of the algorithm under consideration. This work is the extension of the recent work on parameter tuning by Joy et al. (2024) presented at the International Conference on Computational Science (ICCS 2024), and the Firefly Algorithm (FA) is tuned using three different methods: the Monte Carlo method, the Quasi-Monte Carlo method and the Latin Hypercube Sampling. The FA with the tuned parameters is then used to solve a set of six different optimization problems, and the possible effect of parameter setting on the quality of the optimal solutions is  analyzed. Rigorous statistical hypothesis tests have been carried out, including Student's t-tests, F-tests, non-parametric Friedman tests and ANOVA. Results show that the performance of the FA is not influenced by the tuning methods used. In addition, the tuned parameter values are largely independent of the tuning methods used. This indicates that the FA can be flexible and equally effective in solving optimization problems, and any of the three tuning methods can be used to tune its parameters effectively.   \\[10pt]

\noindent {\bf Citation details:} G. Joy, C. Huyck, X.S. Yang, Parameter tuning of the firefly algorithm by three tuning methods: Standard Monte Carlo, quasi-Monte Carlo and latin hypercube sampling methods, {\it Journal of Computational Science}, article 102588 (2025). \\ https://doi.org/10.1016/j.jocs.2025.102588   \\[10pt]


\end{abstract}

\section{Introduction}
Many problems in science and engineering as well as industrial designs
can be formulated as optimization problems with a main objective, subject to multiple nonlinear constraints.  To find the optimal solutions to such optimization problems requires the use of sophisticated optimization algorithms and techniques~\cite{Birattari2009,EibenSmitParametertuning2011,YangHe2019,Yang2020nature}. There are many different techniques for optimization, such as gradient-based methods, gradient-free methods, evolutionary algorithms and nature-inspired metaheuristic algorithms.

A major trend in solving nonlinear optimization problems is to use nature-inspired algorithms, which is especially true in the engineering design and industrial applications. The main advantages of nature-inspired algorithms are that they are effective, flexible and easy to implement. {A recent comprehensive review shows that there are more 540 nature-inspired metaheuristic algorithms in the current literature~\cite{Rajwar2023}, and the literature is still expanding.
The performance of these new algorithms can vary significantly, from the very effective ones to almost useless ones. In addition, the performance of such algorithms may be heavily dependent on their parameter settings, especially those algorithms with multiple parameters. Consequently, fine-tuning of algorithm-dependent parameters is very important to ensure the effectiveness and the proper implementations of such algorithms for their use in real-world applications~\cite{yangStochastictest2010,YangBookNatureInspired2020}.
}

For a given set of optimization problems and a given algorithm to solve them, it is still a challenging task to tune the parameters of the algorithm properly before it can be used to solve many different optimization problems efficiently~\cite{JoyReviewPT2023}. Consider that $A$ represents the algorithm with $m$ parameters $p=(p_1, p_2, ..., p_m)$ to be tuned, the tuned algorithm can be used to find the optimal solution $\x_*$, for a given problem $Q$.
The solution to the problem $Q$ can be obtained through an iterative process, by starting with a random solution $\x_0$. The solution is calculated using the iterative equation,
\be \x^{k+1}=A(\x^k, p, Q). \ee
where $k$ represents the solution vector at iteration $k$. When $k$ is large enough, the optimal solution $\x_*$ might be found.

While most researchers focus on finding the optimal solution $\x_*$, for a given optimization problem $Q$, the use of algorithm $A$ without tuning its parameter is not ideal, unless the problem is a small scale problem or not time consuming. Finding the optimal parameter values $p_*$ can help reduce the computational costs involved, by reducing both the algorithm's running time and improving the efficiency of the algorithm with properly tuned parameter settings.

This paper extends the recent work on parameter tuning using standard Monte Carlo and quasi-Monte Carlo methods~\cite{Geethu2024}, presented at the
International Conference on Computational Science (ICCS 2024).
The possible effect of different tuning methods on the parameters of the Firefly algorithm are evaluated. To study the effect of different tuning methods on parameter settings, the three parameters (i.e., $\theta$ , $\beta$
and $\gamma$)  are tuned using the Monte Carlo (MC), Quasi-Monte Carlo (QMC) methods and Latin Hypercube Sampling (LHS). The difference between the three tuning methods on the parameters of the Firefly algorithm are evaluated using six benchmark functions. The best fitness values obtained for the benchmark functions along with the corresponding parameter values generated from  the three tuning methods are then tested for possible significant differences using Student's t-tests, F-tests, Friedman tests and ANOVA. Therefore, the paper is organized as follows: Section 2 reviews the relevant literature on parameter tuning, followed by the introduction of the FA and three tuning methods. {Section 3 introduces the essential ideas of the FA and the different tuning methods used to tune the parameters of the FA. The tuning methods used are the Monte Carlo (MC) method, Quasi-Monte Carlo (QMC) method, and the Latin Hypercube Sampling (LHS). The standard MC has the slowest convergence rate, whereas both QMC and LHS have a faster convergence rate.
Section 4 presents a diverse set of test benchmarks with parameter setups. Section 5 presents the numerical results with the detailed tests of two hypotheses concerning the mean values, whereas Section 6 carries out more comprehensive tests using F-tests on variances, non-parametric Friedman tests and the analysis of variance (ANOVA). Finally, Section 7 concludes with some discussions for further research. }

\section{Literature Review}
The performance of algorithms with parameters depend on the optimal setting of these algorithm-dependent parameters. However, there is no one unique tuning method that helps in achieving the optimal parameter values. Classifying the different tuning methods in the literature, based only on the classification method used, is a difficult task. There are many studies concerning parameter tuning ~\cite{Calvet2016,Hekmat2019,Joshi2020,Huang2020},
parameter control~\cite{Eiben1999,Tatsis2019,Lacerda2021} and hyper-parameter optimization~\cite{Bergstra2012,Yoo2019}.

{There are different ways to carry out parameter tuning~\cite{SmitEiben2009,JoyReviewPT2023}, though it is not an easy task to fully classify all the different tuning methods. In addition, the structures of the tuning method can also vary.
For a given algorithm and a given tuning method (tuner), the actual tuning structure can be carried out either in sequence or in parallel. In this work, the existing methods can loosely be grouped into ten different categories: }

\begin{itemize}

\item Manual tuning: This brute force method is typically very slow because it tends to try every possible combination in a vast parameter space.

\item Systematic scanning: For given ranges of parameter values, this method tends to scan all possible values in a systematic way. This is also a very slow method.

\item Empirical tuning: This method often starts with a known set of values or ranges based on experience or observations, and then varies the parameters around the known values. Sometimes, a full parametric study may be carried out.

\item Monte Carlo (MC) based methods: MC-based methods are a class of methods that generate random parameter settings from a known distribution, usually a uniform distribution.

\item Tuning by design of experiments (DOE): This approach uses the systematical techniques based design of experiments to initialize or discrete parameter values.

\item Machine learning (ML) based methods: ML-based methods are a class of methods that use machine learning to set and learn new setting of parameters based on the performance of the algorithm.

\item Adaptive and automation methods: The main idea of adaptive and automatic tuning is to use a predefined rule to modify or adjust the parameter values, based on the quality of the solution obtained during the iterations.

\item Self-tuning method: The self-tuning method is to use the algorithm under tuning to tune the parameters of the algorithm~\cite{Yang2013STA}. The parameters are considered as part of the design parameter space, and both the optimal solution and optimal parameter settings are sought simultaneously.

\item Heuristic tuning with parameter control: Heuristic tuning usually uses a heuristic rule to tune the parameters and then varies the parameter values if necessary, based on the quality of the solution. However, there are no known good rules and tuning is by trial and error.

\item Other tuning methods: {There are other methods that are not put into the above categories. This includes multi-objective approach for parameter tuning, sequential optimization, fuzzy method, Baysian optimization~\cite{Garnett2023bayesian} and others. }
\end{itemize}	

\begin{figure}
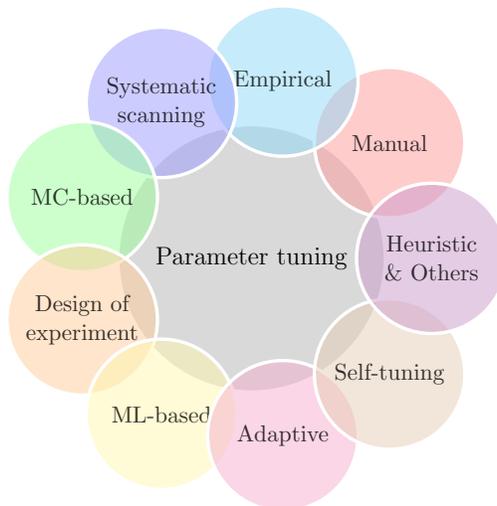
 \begin{center}
\scalebox{0.8}{
\smartdiagram[bubble diagram]{Parameter tuning,
Manual, Empirical, Systematic \\ scanning,  MC-based, Design of \\ experiment, ML-based,
Adaptive, Self-tuning, Heuristic \\ \& Others}
}
\caption{Main methods for parameter tuning. }
\label{Fig-tuning-100}
\end{center} \end{figure}

The main methods are summarized in Fig.~\ref{Fig-tuning-100}. Despite extensive research, there are still many unresolved issues and open problems concerning parameter tuning.

To be more specific, there are three open problems:

\begin{enumerate}

\item {\it Non-universality}. For a given algorithm and a set of optimization problems, it is  unclear if tuned parameters of a given algorithm can be applied to solve other problems with the same efficiency. In other words, the main question is that: are parameter settings problem-specific and algorithm-specific?

\item {\it High computational efforts}. Parameter tuning is a time-consuming task. What methods should be used to minimize the overall computational efforts?

\item {\it Lack of theoretical insights}. Though there are quite a few tuning methods in the current literature, most such methods are heuristic approaches or statistical methods. There are no theoretical guidelines for the effective ways of tuning parameters. Some theoretical analyses are desperately needed to gain some understanding of the search mechanisms and optimal tuning conditions.
\end{enumerate}

In addition to the above open problems, {this work intends to answer two main questions concerning parameter tuning:}
\begin{itemize}
\item {For a given algorithm to solve a given set of optimization problems, is the solution quality (in terms of the optimal objective values) affected by the tuning method used?}

\item {For a given algorithm, are the tuned parameter values of the algorithm affected by different tuning methods? }

\end{itemize}

{In this paper, the FA is used to evaluate the possible effects of parameter settings on the performance of the algorithm for a set of different benchmarks. Comprehensive analyses of the results will be carried out related to above two key questions. }

\section{Tuning the Firefly Algorithm by MC, QMC and LHS}

\begin{figure}
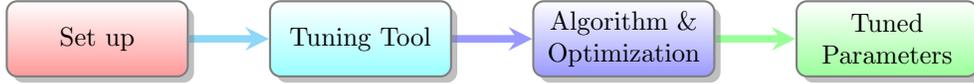

\begin{center}
\smartdiagramset{back arrow disabled=true,text width=2.15cm,
module x sep=3.5cm}
\smartdiagram[flow diagram:horizontal]{Set up, Tuning Tool,
Algorithm \& \\ Optimization, Tuned Parameters}
\caption{Main steps of parameter tuning for a given algorithm.}
\label{Fig-pt-100}
\end{center}
\end{figure}

The main steps of the parameter tuning in this paper can be schematically represented in Fig.~\ref{Fig-pt-100} where the setup step is to ensure both the algorithm and the optimization problem to be solved are properly implemented. The tuning tool (i.e., tuner) is used to initialize the parameters to be used in the algorithm, which in turn will be used to solve the optimization. Based on the quality of the solutions obtained from different settings, the best parameter values and best objective values are selected and stored as the results.

Before the details of tuning parameters using the Monte Carlo (MC), Quasi-Monte Carlo (QMC) and Latin Hypercube Sampling (LHS) methods are discussed, the main idea of FA and its parameters are outlined.

\subsection{Firefly Algorithm}

The Firefly Algorithm (FA) is a nature-inspired algorithm for multimodal optimization, developed by Xin-She Yang in 2008, based on the flashing characteristics and flight patterns of tropical fireflies~\cite{yangStochastictest2010}. Due to its simplicity,
FA has been applied to a wide range of applications, including mechanical design problems~\cite{Bayka2015,Erdal2017}, color image segmentation~\cite{HeHuang2017}, power system optimization~\cite{Singh2017}, vehicle routing problems~\cite{Goel2018}, security enhancement in communications~\cite{Vien2019}, optimal transmit beam forming~\cite{TuanLe2024} and engineering design optimization~\cite{El-Shor2022}.
	
The solution vector $\x$ to an optimization problem is encoded
as the locations of fireflies. Therefore, two solution vectors correspond to the two locations of two fireflies $i$ and $j$  at $\x_i$ and $\x_j$, respectively.  The algorithmic equation of the FA is to update the solution vectors by
\begin{equation}
\x_{i}^{t+1} = \x_{i}^{t} + \beta e^{-\gamma r_{ij}^2}
	(\x_{j} - \x_{i}) + \alpha \epsilon_{i}^{t}, \label{Position-Update}
	\end{equation}
where $\epsilon_i^t$ represents a vector of random numbers that are usually drawn from a normal distribution $N(0,1)$.  The distance $r_{ij}$ between two solution vectors, $\x_i$ and $\x_j$, is given by the Euclidean distance or $L_2$-norm
\begin{equation}
\quad r_{ij} = \lVert \x_{i}^{t} - \x_{j}^{t} \rVert.
\end{equation}
In the FA,  there are three parameters to be tuned, and they are the attractiveness $\beta$, the scaling parameter $\gamma$ and the perturbation strength parameter $\alpha$. In most FA implementations, parameter $\alpha$ is further rewritten as
\begin{equation}
\alpha=\alpha_0 \theta^t,
\end{equation}
where $\alpha_0$ is its initial value and can be set as $\alpha_0=1$. The iteration counter $t$ is the pseudo-time. Instead of tuning $\alpha$,  $0<\theta<1$ is the parameter to be tuned in this paper.

\begin{figure}
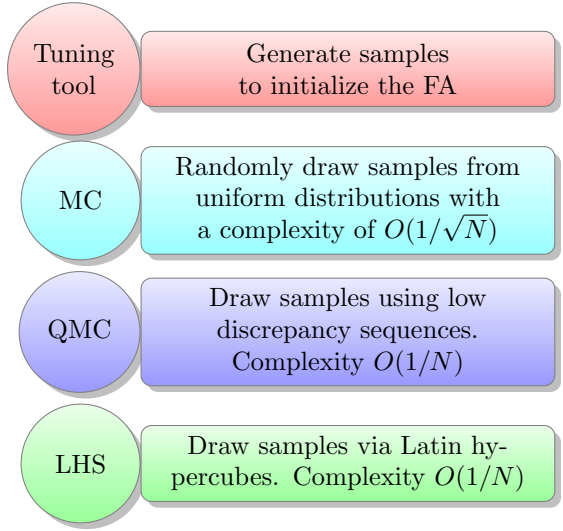
 \begin{center}
\smartdiagram[descriptive diagram]{
{Tuning tool, Generate samples to initialize the FA},
{MC, Randomly draw samples from uniform distributions with a complexity of $O(1/\sqrt{N})$},
{QMC, Draw samples using low discrepancy sequences. Complexity $O(1/N)$},
{LHS, Draw samples via Latin hypercubes. Complexity $O(1/N)$},
}
\caption{{Three different tuning methods. }}
\label{Fig-three-100}
\end{center}\end{figure}

The tuning tool is a tuner, which can be any parameter tuning method. In this study, three different tuning methods are used, including the standard Monte Caro (MC), Quasi-Monte Carlo (QMC) and Latin Hypercube Sampling (LHS) methods (see Fig.~\ref{Fig-three-100}) where $N$ is the sample size.  The main ideas of the three tuning methods are briefly explained below.

\subsection{Monte Carlo  Method}

{The Monte Carlo method is a commonly used method for many simulation tasks, including engineering simulation, parameter tuning, climate modelling and stastical sampling. It works more effectively when compared to the manual or brute force tuning method. The MC method has a statistical foundation and its errors decrease as $O(1/\sqrt{N})$, where $N$ is the number of samples. In this study, the MC  method is used to randomly initialize the parameters of the FA. The MC generates uniformly distributed pseudo-random numbers from which discrete samples within a specified range are drawn and then used as the initial values of parameters $\theta$, $\beta$ and $\gamma$.}

\subsection{Quasi-Monte Carlo Method}

{The QMC method requires fewer samples than the MC method because its errors decrease with sample size $N$ as $O(1/N)$, which also leads to a faster convergence rate in practice. The QMC generates quasi-random numbers between the interval $0$ and $1$ using low-discrepancy sequences such as the Van der Corput sequence, the Halton sequence and the Sobol sequence. In this study, the standard Matlab implementation of the Sobol sequence with digital shifts and affine scramble has been used. }

\subsection{Latin Hypercube Sampling}

{The LHS method is more effective when used in tasks involving huge computational cost, as it uses fewer samples than the MC method. It is a stratified sampling method based on Latin squares. The LHS method reduces the computational time required due to the lower number of samples generated and its errors decrease as $O(1/N)$. }

\section{Experiment Setup and Test Benchmarks}

\subsection{Experimental Setup for FA Parameters}
To test the possible effects of different tuning methods on the performance of the FA, the parameters of the FA are generated using the MC, QMC and LHS tuning methods. The best fitness values obtained for each benchmark function using these parameter values are then tested for significant differences using different statistical tests.

In the standard FA, there are three parameters $\theta$, $\beta$ and $\gamma$ to be tuned. These parameters of the FA can typically take the following values:

\begin{itemize}
\item The size of the population in the FA: $n = 20$ to $40$ (up to $100$ if necessary).
	
\item $\beta = 0.1$ to $1$, though $\beta=1$ is a typical value.

\item $\gamma = 0.01$ to $10$, though typically $\gamma = 0.1$ to $1$.
	
\item  $\alpha_0 = 1$, and $\theta = 0.9$ to $0.99$ (its typical value is $\theta = 0.97$). Here, $\alpha = \alpha_0  \theta^{t}$.
	
\item Maximum number of iterations: $t_{\max} = 100$ to $1000$.
\end{itemize}
For simplicity in this study for all three tuning methods, the ranges of all the relevant parameter values are summarized in Table~\ref{table-100}.

{All the simulations have been performed on a computer using MATLAB R2023a on Windows 11 with a hardware configuration of multi-core CPUs of 2.40GHz and 8 GB RAM.}

\begin{table}
\begin{center}	
\caption{Experimental setting for MC, QMC and LHS Simulations}
\label{table-100}
\begin{tabular}{|c|c|c|c|}
\hline
Initialization Values & MC & QMC & LHS \\
\hline
Size of the population  &  20 & 20 & 20 \\
\# of  runs &  10 & 10 & 10 \\
\# of iterations & 1000 & 1000 & 1000 \\ \hline
Parameter ranges of & & &  \\
$\theta$ & [0.9,  1.0] & [0.9, 1.0] & [0.9, 1.0]\\
$\beta$ & [0, 1] & [0, 1] & [0, 1] \\
$\gamma$  & [0.1,  2.5] & [0.1,  2.5] & [0.1, 2.5]  \\
\hline
\end{tabular}
\end{center}
\end{table}

In addition, 10 independent runs have been carried out for each method, and each run has a different parameter setting generated by the chosen method.  Within each independent run with the same parameter values, the population of the firefly algorithm has been initialized randomly, and the FA has been called 50 times to solve the same optimization problem.  After the 50 calls of the FA, the best result is recorded as the result for each method's independent run. Therefore, there are 10*50=500 simulation realizations in the numerical experiments in this paper.

\subsection{Benchmark Functions}

The efficiency and reliability of algorithms are generally tested using benchmark or test functions. There are many different benchmark collections such as the CEC suites, review articles by Jamil and Yang~\cite{JamilYang2013}, and online collection like GAMS. The choice of test benchmarks should be sufficiently diverse in terms of mathematical properties such as modality, convexity, separability, dimensionality, linearity and non linearity. Unlike functions, which are generally unconstrained, real world benchmark problems have complex constraints. In this study six different benchmarks are used. The first three functions, which are the Sphere, Rosenbrock and Ackley function, have their optimal objectives at $f_{\min}=0$, whereas the other three functions which are Trid, Spring design and Truss design have non-zero objectives. These benchmarks are outlined below:

\begin{enumerate}

\item The simplest test function is the \emph{sphere function}, which is both convex and separable in the form
\begin{equation}
	f_1(\x) = \sum_{i=1}^D x_{i}^{2}, \quad \x \in \mathbb{R}^D,
\end{equation}
with simple limits of
\begin{equation}
\quad -10 \leq x_i \leq 10, \quad i=1,2,..., D.
\end{equation}
The global minimum of this function is located at $\x^* = (0, \ldots, 0)$ with $f_{\min} = 0$.

\item The \emph{Rosenbrock function} is a nonlinear, nonconvex,  benchmark~\cite{JamilYang2013} in the $D$-dimensional space
\begin{equation}
f_2(\x) = (1-x_1)^2+ \sum_{i=1}^{D-1} \left[ 100\left(x_{i+1}-x_i^2\right)^2  \right], \quad \x \in \mathbb{R}^D,
\end{equation}
with simple limits of
\begin{equation}
-30 \le x_i \le 30, \quad i=1,2,..., D.
\end{equation}
Its global minimum $f_{\min}=0$ is located at $\x^* = (1, \ldots, 1)$.

\item  The \emph{Ackley function} is a multimodal, nonconvex function
\be f_3(\x)=-20 \exp\left[-0.2 \sqrt{\frac{1}{D} \sum_{i=1}^D x_i^2}\right]
-\exp\left[\frac{1}{D} \sum_{i=1}^D \cos(2 \pi x_i)\right] +20 +e, \ee
where
\be x_i \in [-32.768, 32.768], \quad i=1,2,..., D. \ee
Its global minimum $f_{\min}=0$ is located at $(0,0,...,0)$.

\item The \emph{Trid function} is a non-separable function with non-zero optimal objective
\be f_4(\x)=\sum_{i=1}^D (x_i-1)^2 - \sum_{i=2}^D x_i x_{i-1}, \quad
x_i \in [-D^2, D^2]. \ee
The location of its global minimum depends on the dimensionality of the function with the global minimum
\be f_{\min}=-\frac{D(D+4)(D-1)}{6}, \quad x_i=i(D+1-i), \quad (i=1,2,...,D). \ee
In case of $D=4$, the optimality $f_{\min}=-16$ occurs at $\x_*=(4, 6, 6, 4)$.

\item The \emph{spring design} is a design benchmark in engineering applications. It has three decision variables, subject to some highly nonlinear constraints~\cite{YangBookNatureInspired2020}.
\begin{equation}
	\textrm{Minimize } \; f_5(\x) = (2 + x_3) x_1^2 x_2,
\end{equation}
subject to
\begin{align}	
	& g_1(x) =	1 - \frac{x_2^{3}x_3}{71785x_1^{4}} \leq 0, \nonumber \\
	& g_2(x) = \frac{4x_2^2 - x_1x_2}{12566(x_2x_1^{3} - x_1^{4})} + \frac{1}{5108x_1^{2}} - 1 \leq 0, \nonumber \\
	& g_3(x) = 1 - \frac{140.45x_1}{x_2^{2}x_3} \leq 0, \nonumber \\
	& g_4(x) = \frac {x_1+x_2}{1.5} \leq 0. \nonumber
\end{align}
The simple bounds for design variables are
\begin{equation}
0.05 \le x_1 \le 2.0, \quad 0.25 \le x_2 \le 1.3,
\quad 2.0 \le x_3 \le 15.0.
\end{equation}
Though the true optimal solution is still unknown,
the best solution in the literature~\cite{Cagnina2008} is
\begin{equation}
\x_*=[0.051690, \; 0.356750, \; 11.287126],
\quad f_{\min}(\x_*)=0.012665.
\end{equation}
All the constraints in this optimization problem are handled by using the standard penalty method.

\item The \emph{truss design} system, also called the three-bar truss design problem, has two design variables as cross-section areas. The objective is to minimize
\be f_6(\x)=100 (2 \sqrt{2} x_1 + x_2), \ee
subject to three stress constraints
\begin{align}
g_1(\x) & = \frac{(\sqrt{2} x_1 + x_2) P }{\sqrt{2} x_1^2 + 2 x_1 x_2}-\sigma \le 0, \\
g_2(\x) & = \frac{x_2 P}{\sqrt{2} x_1^2 + 2 x_1 x_2} -\sigma \le 0, \\
g_3(\x) & = \frac{P}{x_1 + \sqrt{2} x_2} -\sigma \le 0.
\end{align}

{In addition, the stress limit is $\sigma=2000$ N/cm$^2$, and the load is $P=2000$ N.
Though most formulations in the literature used $x_1, x_2 \in (0, 1]$, however, both $x_1$ and $x_2$ cannot be zero because they are physical quantities. Thus, to ensure the proper requirement of the physical quantities, a very small lower bound of $0.001$ is imposed on both $x_1$ and $x_2$, and the actual bounds are $x_1, x_2 \in [0.001, 1]$. This will ensure all the design requirements can be met in the implementations of this optimization task.}

Though the true optimal solution is unknown, the best solution in the literature~\cite{Bekdas2018} is
\be f_{\min}=263.8958, \quad \x_*=(0.78853, 0.40866). \ee

\end{enumerate}
In summary, the first three benchmark functions (i.e., the sphere, Rosenbrock and Ackley function) have the optimal value $f_{\min}(\x^*)$ at 0, whereas the remaining three benchmark functions (i.e., the Trid, spring design and three-bar truss design) have non-zero optimal values. In addition, some of the benchmark functions selected are convex (e.g. sphere), most of these benchmarks are non-convex. The spring design and the truss design considered for this study are constrained problems, subject to nonlinear constraints. These constraints are handled properly by using the penalty method.

{For the constrained optimization benchmarks, all the constraints are handled by the standard penalty method~\cite{Yang2020nature}
\be \Pi(\x)=f(\x) + \lambda \sum_{j=1}^K \max\{0, g_j(\x)\}, \ee
where $K$ is the number of inequality constraints $g_j(\x)$, and $f(\x)$ is the original objective function.
Here, parameter $\lambda$ is the penalty coefficient, which is set to $\lambda=1000$ in the implementations.
}

\section{Simulation Results and Hypothesis testing}
To investigate the possible effect of the three different tuning methods on the Firefly algorithm, the best fitness values obtained for the six benchmark functions along with the corresponding parameter values are tested using two hypotheses. The results obtained from the ten simulation runs of MC, QMC and LHS each are evaluated using two hypotheses, where the first hypothesis focus on evaluating the effect of tuning methods on the fitness values obtained and the second hypothesis  emphasizes the evaluation of the possible effect of tuning methods on the parameter values of the FA.

The two hypotheses to be tested are as follows
\begin{enumerate}

\item[] { {\bf Hypothesis H1}: For a given optimization problem to be solved by a given algorithm, parameter tuning methods (MC, QMC or LHS) have no significant effect on the objective values obtained.  \\    }

\item[] { {\bf Hypothesis H2}: For a given algorithm to solve a set of optimizatin problems, its performance and tuned parameter values are not affected by the parameter tuning method used. \\
}
\end{enumerate}

\subsection{Summary of Numerical Experiments}
The best fitness values along with the corresponding parameter values obtained for each of the six benchmark functions, using the three tuning methods MC, QMC and LHS are investigated in this section. The fitness value listed in each table is the best fitness value identified from the 10 runs of each simulation method. These optimal fitness values are calculated using the parameter values generated using MC, QMC and LHS methods. Although summary statistics indicate the fitness values obtained from all three tuning methods to be of the same order, for all six benchmark functions, this may only be determined using additional statistical tests.

The optimal objective values found by the FA for all the six benchmarks from ten runs of MC are summarized in Table~\ref{Tab-MC-100}.
\begin{table}[ht]
\begin{center}
\caption{{Objective values by the MC method.}}
\label{Tab-MC-100}
\begin{tabular}{c|cccccc}
\hline
Run & $f_1$ & $f_2$ & $f_3$ & $f_4$ & $f_5$ & $f_6$ \\
\hline
1 & 4.4697e-292  &  1.6512e-03  &  4.1628e-02  &  -15.941  &  1.9458e-02  &  263.8958 \\
2 & 1.9858e-203  &  9.6323e-03  &  0.0000e+00  &  -15.941  &  2.2446e-02  &  263.8958 \\
3 & 8.7906e-120  &  5.9862e-02  &  2.2204e-15  &  -16.000  &  2.2504e-02  &  263.8958 \\
4 & 0.0000e+00  &  3.4996e-04  &  0.0000e+00  &  -16.000  &  1.5534e-02  &  263.8980 \\
5 & 7.9140e-04  &  1.4130e-03  &  0.0000e+00  &  -15.828  &  1.4992e-02  &  263.8959 \\
6 & 0.0000e+00  &  6.8866e-03  &  0.0000e+00  &  -15.657  &  1.5871e-02  &  263.8958 \\
7 & 1.4218e-292  &  2.6163e-02  &  0.0000e+00  &  -15.986  &  1.8961e-02  &  263.8959 \\
8 & 1.1631e-292  &  8.8692e-02  &  1.4345e+00  &  -16.000  &  1.4524e-02  &  263.8958 \\
9 & 0.0000e+00  &  1.2839e-03  &  0.0000e+00  &  -15.862  &  2.0764e-02  &  263.8959 \\
10 & 0.0000e+00  &  3.0315e-03  &  2.2204e-15  &  -16.000  &  1.9090e-02  &  263.8958 \\ \hline
Mean & 7.9140e-5 &  1.9897e-02 & 1.4761e-01 & -15.922 & 1.8414e-02 &
263.8961 \\ \hline
$\sigma$ & 2.5026e-04 & 3.0425e-02 & 4.5236e-01 & 0.1115 & 3.0177e-02 &
0.0007 \\
\hline
\end{tabular}
\end{center}
\end{table}

Similarly, all the results for six benchmarks after ten runs using QMC are summarized in Table~\ref{Tab-QMC-100} and all fitness values obtained from ten runs of LHS using FA are summarized in table ~\ref{Tab-LHR-100}.

\begin{table}[ht]
\begin{center}
\caption{{Objective values for six benchmarks by the QMC method.}}
\label{Tab-QMC-100}
\begin{tabular}{c|cccccc}
\hline
Run & $f_1$ & $f_2$ & $f_3$ & $f_4$ & $f_5$ & $f_6$ \\
\hline
1 & 0.0000e+00  &  8.3758e-03  &  9.6898e-01  &  -16.000  &  1.4092e-02  &  263.8958 \\
2 & 8.1104e-142  &  2.0797e-03  &  0.0000e+00  &  -16.000  &  1.5997e-02  &  263.8959 \\
3 & 1.0233e-17  &  1.2885e-04  &  2.8238e-01  &  -16.000  &  1.9251e-02  &  263.8958 \\
4 & 0.0000e+00  &  1.6545e-05  &  0.0000e+00  &  -15.730  &  1.6024e-02  &  263.8958 \\
5 & 4.2661e-04  &  3.6217e-01  &  1.1298e+00  &  -16.000  &  2.1156e-02  &  263.8958 \\
6 & 2.7031e-305  &  1.2378e-03  &  0.0000e+00  &  -15.988  &  1.7453e-02  &  263.8960 \\
7 & 0.0000e+00  &  1.3343e-02  &  2.2204e-15  &  -16.000  &  1.7496e-02  &  263.8958 \\
8 & 0.0000e+00  &  3.1957e-02  &  2.2204e-15  &  -15.915  &  1.6202e-02  &  263.8958 \\
9 & 7.4649e-14  &  1.0026e-02  &  3.2142e-01  &  -16.000  &  1.8144e-02  &  263.8959 \\
10 & 1.1193e-188  &  7.7233e-03  &  0.0000e+00  &  -16.000  &  1.4354e-02  &  263.8980 \\ \hline
Mean & 4.26610e-5 &  4.3706e-02 & 2.7026e-01 & -15.963 & 1.7017e-02 &
263.8961 \\ \hline
$\sigma$ & 1.3491e-04 & 1.1229e-01 & 4.3051e-01 & 0.086 & 2.1641e-02 &
0.0007 \\
\hline
\end{tabular}
\end{center}
\end{table}

\begin{table}[ht]
\begin{center}
\caption{{Objective values by the LHS.}}
\label{Tab-LHR-100}
\begin{tabular}{c|cccccc}
\hline
Run & $f_1$ & $f_2$ & $f_3$ & $f_4$ & $f_5$ & $f_6$ \\
\hline
1 & 0.0000e+00  &  1.9713e-01  &  0.0000e+00  &  -15.996  &  1.4569e-02  &  263.8958 \\
2 & 4.7252e-07  &  4.7954e-02  &  0.0000e+00  &  -15.995  &  1.7682e-02  &  263.8961 \\
3 & 1.9924e-194  &  8.8756e-03  &  0.0000e+00  &  -16.000  &  1.3736e-02  &  263.8958 \\
4 &  6.1393e-04  &  1.7592e-01  &  1.5460e-07  &  -15.834  &  1.8583e-02  &  263.8958 \\
5 & 0.0000e+00  &  1.8493e-01  &  0.0000e+00  &  -15.792  &  1.4276e-02  &  263.8958 \\
6 & 2.1416e-165  &  3.7766e-02  &  0.0000e+00  &  -16.000  &  2.1030e-02  &  263.8958 \\
7 & 0.0000e+00  &  4.3180e-04  &  0.0000e+00  &  -15.942  &  1.4930e-02  &  263.8959 \\
8 & 0.0000e+00  &  1.7003e-03  &  0.0000e+00  &  -16.000  &  1.2939e-02  &  263.8958 \\
9 & 9.4408e-05  &  1.4994e-01  &  2.2204e-15  &  -16.000  &  2.2164e-02  &  263.8958 \\
10 & 3.8103e-04  &  8.9662e-04  &  3.6561e-02  &  -16.000  &  2.5559e-02  &  263.8958 \\
\hline
Mean & 1.0898e-4 &  8.0554e-02 & 3.6561e-03 & -15.956 & 1.7547e-02 &
263.8958 \\ \hline
$\sigma$ & 2.1402e-04 & 8.5249e-02 & 1.1562e-01 & 0.078 & 4.2283e-03 &
0.0001 \\
\hline
\end{tabular}
\end{center}
\end{table}

\subsection{Testing the First Hypothesis}

{There are many different methods for testing hypotheses, and comprehensive tests using different methods will be carried out in this work. The first method to be used here for testing the two hypotheses is the paired Student t-tests. The primary reason to use t-tests is that such tests are valid for testing the differences in means for small sample sizes (usually under 30). This is because the t-test assumes the underlying data to be normally distributed, even when the number of samples is small~\cite{Mooi2018hypothesis,Geraldttest2018}.
}

The main test statistic for two-sample tests is
\be t=\frac{\bar x -\bar y}{\sqrt{S_x^2/N_x + S_y^2/N_y}}, \ee
where $\bar x$ and $\bar y$ are the sample means of $x_i \, (i=1,2,..., N_x)$ and $y_j \, (j=1,2,..., N_y)$, respectively.  $S_x^2$ and $S_y^2$ are their
corresponding sample variances. In case of equal sample sizes, $N=N_x=N_y$
can be used.

\begin{table}[ht]
\begin{center}
\caption{{The p-values of paired Student t-tests.}}
\label{Tab-p-t-100}
\begin{tabular}{|l|ccc|}
\hline
Function & MC vs QMC & MC vs LHS & QMC vs LHS \\
\hline
Sphere $f_1$ & 0.6897  &   0.7777  &   0.4180 \\
Robsenbrock $f_2$ & 0.5247  &   0.0482  &   0.4194 \\
Ackley $f_3$ & 0.5423  &   0.3277  &   0.0660 \\
Trid $f_4$ & 0.3605  &   0.4344  &   0.8427 \\
Spring $f_5$ & 0.2495  &   0.6038  &   0.7283 \\
Truss $f_6$ & 0.9744 & 0.3510  & 0.3280 \\
\hline
\end{tabular}
\end{center}
\end{table}

The p-values for paired Student t-tests are summarized in Table~\ref{Tab-p-t-100}. {As seen in the table~\ref{Tab-p-t-100}, for the sphere function $f_1$,  the paired t-tests show that the p-vales are 0.6897, 0.7777, and 0.4180 for MC versus QMC, MC versus LHS, and QMC versus LHS, respectively.
All p-values are greater than the critical value of 0.05, thus the null hypothesis cannot be rejected.
That is, there are no significant differences in terms of the objective values found by three tuning methods.

For the Rosenbrock function $f_2$, the p-values are 0.5247, 0.0482, and 0.4194 for MC versus QMC, MC versus LHS, and QMC versus LHS, respectively. It seems that there is a marginal difference between MC and LHS, though the p-value is quite close to the critical value 0.05. To make sure, further tests will be carried out later in this section. For the Ackley function $f_3$, the p-values are 0.5423, 0.3277, and 0.0660. Thus, there are no significant differences in the objective values, as the p-values obtained from the paired t-tests are all greater than $0.05$.
For the Trid function $f_4$, the p-values obtained from paired t-tests are 0.3605, 0.4344, and 0.8427, which are all above 0.05. Thus, no significant differences are observed in the fitness value obtained using the three different tuning methods.

For the spring design $f_5$, the p-values obtained are 0.2495, 0.6038, and 0.7283, respectively. This further indicates the insensitivity of fitness values towards the different tuning methods. For the truss system $f_6$, the p-values obtained from the paired t-tests of MC with QMC, MC with LHS and QMC with LHS are 0.9744, 0.3510 and 0.3280, respectively. The higher p-values suggest that the tuning methods have no significant effect on the fitness values obtained.

From the p-values calculated in the above tests, all tests indicate that no significant differences are observed in the objective values obtained by the FA with three different tuning methods. }

\subsection{Varying Sample Sizes}
For the Rosenbrock function, the initial sample size is $10$, which gives one of the p-values as 0.0482. To make sure the sample size is large enough, 30 independent runs have been carried out and the results of the objective values are summarized in
Table~\ref{Tab-new-Rosen-100}.

\begin{table}[ht]
\begin{center}
\caption{Results from 30 independent runs for the Rosenbrock function.}
\label{Tab-new-Rosen-100}
\begin{tabular}{|l|r|r|r|}
\hline
Run \# & MC & QMC & LHS \\
\hline
1 & 1.6512e-03  &  8.3758e-03  &  1.9713e-01 \\
2 & 9.6323e-03  &  2.0797e-03  &  4.7954e-02 \\
3 & 5.9862e-02  &  1.2885e-04  &  8.8756e-03 \\
4 & 3.4996e-04  &  1.6545e-05  &  1.7592e-01 \\
5 & 1.4130e-03  &  3.6217e-01  &  1.8493e-01 \\
6 & 6.8866e-03  &  1.2378e-03  &  3.7766e-02 \\
7 & 2.6163e-02  &  1.3343e-02  &  4.3180e-04 \\
8 & 8.8692e-02  &  3.1957e-02  &  1.7003e-03 \\
9 & 1.2839e-03  &  1.0026e-02  &  1.4994e-01 \\
10 & 3.0315e-03  &  7.7233e-03  &  8.9662e-04 \\
11 & 1.8843e-01  &  4.4617e-04  &  1.2139e-02 \\
12 & 2.5424e-03  &  4.0468e-02  &  2.1478e-02 \\
13 & 4.7216e-02  &  2.1764e-01  &  3.3757e-02 \\
14 & 4.0512e-02  &  1.1109e-03  &  5.6404e-03 \\
15 & 3.6326e-04  &  5.0980e-02  &  7.2126e-03 \\
16 & 4.8076e-08  &  1.5776e-03  &  1.9084e-07 \\
17 & 1.4043e-02  &  8.7533e-04  &  2.0073e-05 \\
18 & 2.2634e-02  &  6.8443e-03  &  1.9291e-02 \\
19 & 4.2940e-05  &  7.9083e-03  &  8.0240e-02 \\
20 & 7.8361e-05  &  1.5354e-02  &  1.5894e-01 \\
21 & 1.2593e-02  &  1.6331e-03  &  1.0276e-03 \\
22 & 1.1930e-05  &  3.3622e-05  &  4.1166e-02 \\
23 & 5.4078e-02  &  6.9899e-02  &  5.4658e-02 \\
24 & 2.4579e-02  &  4.4231e-02  &  5.1685e-01 \\
25 & 2.5262e-02  &  2.1813e-03  &  5.6887e-03 \\
26 & 9.7552e-03  &  4.1163e-02  &  2.2985e-02 \\
27 & 1.1581e-04  &  2.5131e-03  &  2.3202e-03 \\
28 & 8.4856e-03  &  3.3293e-02  &  8.5775e-05 \\
29 & 5.0365e-02  &  5.7525e-02  &  1.5274e-02 \\
30 & 2.4584e-03  &  7.0070e-09  &  1.2288e-07 \\
\hline
\end{tabular}
\end{center}
\end{table}

{Using the simulation results summarized in Table~\ref{Tab-new-Rosen-100}, the p-values of the paired t-tests
are $0.4762$ for MC versus QMC, $0.0807$ for MC versus LHS, and $0.2381$ for QMC and LHS,
respectively. All these p-values are greater than 0.05. Therefore, there are no significant differences in the objective values obtained by the three tuning methods.

Similarly, a set of 30 independent runs have also been carried out for the Ackley function.
The p-values from the paired t-tests are $0.2724$ for MC versus QMC, $0.4518$ for MC versus LHS,
and $0.6000$ for QMC versus LHS, respectively. As all p-values are greater than 0.05, this further confirms
that no significant differences are observed in the objective values obtained by the three tuning methods. These conclusions are consistent with the earlier observations and results. }

\subsection{Testing the Second Hypothesis}
The second hypothesis evaluates the effect of different tuning methods on the parameters of FA. Student's t-tests are used to conduct pair-wise comparison of the parameter values of $\theta$, $\beta$ and $\gamma$. The values of all three parameters, for all the six benchmark functions corresponding to the best fitness values calculated, are listed in tables below. The best values for
$\theta$ for all the benchmarks using MC, QMC and LHS are summarized in Table~\ref{Tab-theta-100}.
\begin{table}[ht]
\begin{center}
\caption{{The best $\theta$ values obtained by MC, QMC and LHS.}}
\label{Tab-theta-100}
\begin{tabular}{|l|rrr|}
\hline
Function & MC & QMC & LHS \\
\hline
$f_1$ &  0.2607  &   0.2312  &   0.2352 \\
$f_2$ &  0.4755  &   0.6168  &   0.6704 \\
$f_3$ & 0.7673  &   0.8504  &   0.2352 \\
$f_4$ & 0.8384  &   0.6608  &   0.6384 \\
$f_5$ & 0.6299  &   0.2312  &   0.4624 \\
$f_6$ & 0.1610  &   0.8860  &   0.8448 \\
\hline
mean & 0.5221 & 0.5794 & 0.5144 \\
$\sigma$ & 0.2730 & 0.2891 & 0.2480 \\ \hline
\end{tabular}
\end{center}
\end{table}

\begin{table}[ht]
\begin{center}
\caption{{The best $\beta$ values obtained by MC, QMC and LHS.}}
\label{Tab-beta-100}
\begin{tabular}{|l|rrr|}
\hline
Function & MC & QMC & LHS \\
\hline
$f_1$ & 0.3021  &   0.7118  &   0.5950 \\
$f_2$ & 0.7676  &   0.9364  &   0.9590 \\
$f_3$ &  0.9093  &   0.3635  &   0.5950 \\
$f_4$ & 0.4100  &   0.2821  &   0.9910 \\
$f_5$ & 0.1439  &   0.7118  &   0.6530 \\
$f_6$ & 0.5362  &  0.3911  &   0.5230 \\
\hline
mean     & 0.5115  & 0.5661  & 0.7193 \\
$\sigma$ & 0.2876  & 0.2577  & 0.2025 \\
\hline
\end{tabular}
\end{center}
\end{table}

The best $\beta$ values, obtained by three tuning methods for
all benchmarks, are summarized in Table~\ref{Tab-beta-100}.
In addition, the best $\gamma$ values, obtained by three tuning methods for all the benchmarks, are summarized in Table~\ref{Tab-gamma-100}.

\begin{table}[ht]
\begin{center}
\caption{{The best $\gamma$ values obtained by MC, QMC and LHS.}}
\label{Tab-gamma-100}
\begin{tabular}{|l|rrr|}
\hline
Function & MC & QMC & LHS \\
\hline
$f_1$ & 0.7139  &   0.8224  &   1.4225 \\
$f_2$ & 1.2122  &   1.8378  &   1.0085 \\
$f_3$ & 0.5763  &   1.2665  &   1.4225 \\
$f_4$ & 1.8062  &   0.8983  &   0.5615 \\
$f_5$ & 1.7284  &   0.8224  &   0.6995 \\
$f_6$ & 1.2579  &   0.7718  &   0.7745 \\
\hline
mean & 1.2158  & 1.0699 & 0.9765 \\
$\sigma$ & 0.5048 & 0.4169 & 0.3746 \\
\hline
\end{tabular}
\end{center}
\end{table}

\begin{figure} \centering
\includegraphics[height=2in,width=1.5in]{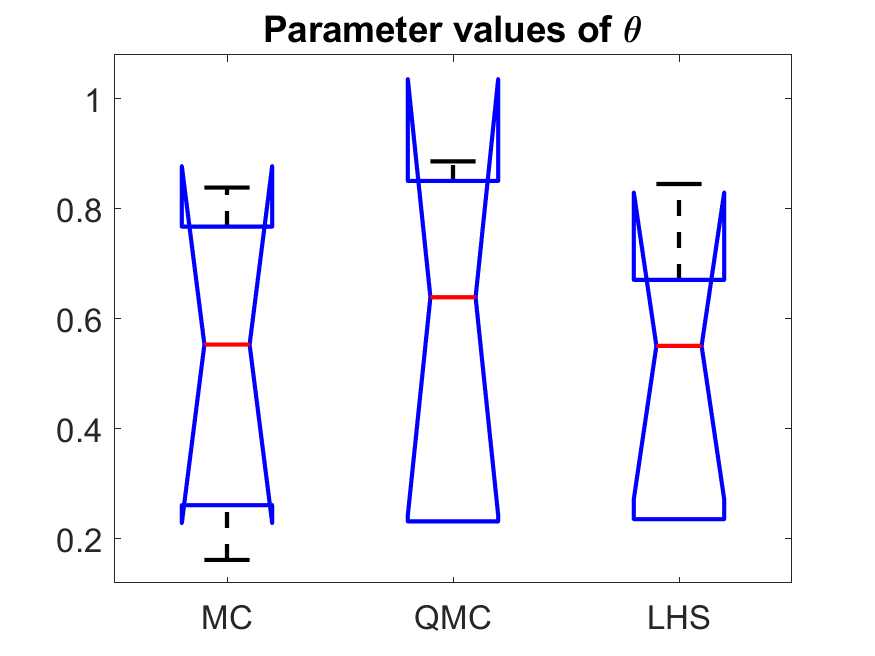}
\includegraphics[height=2in,width=1.5in]{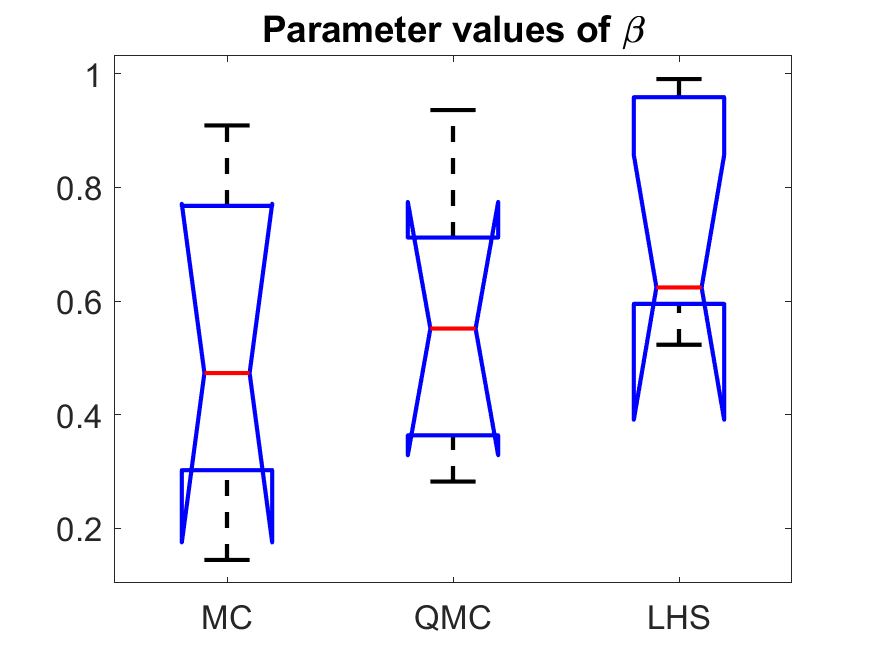}
\includegraphics[height=2in,width=1.5in]{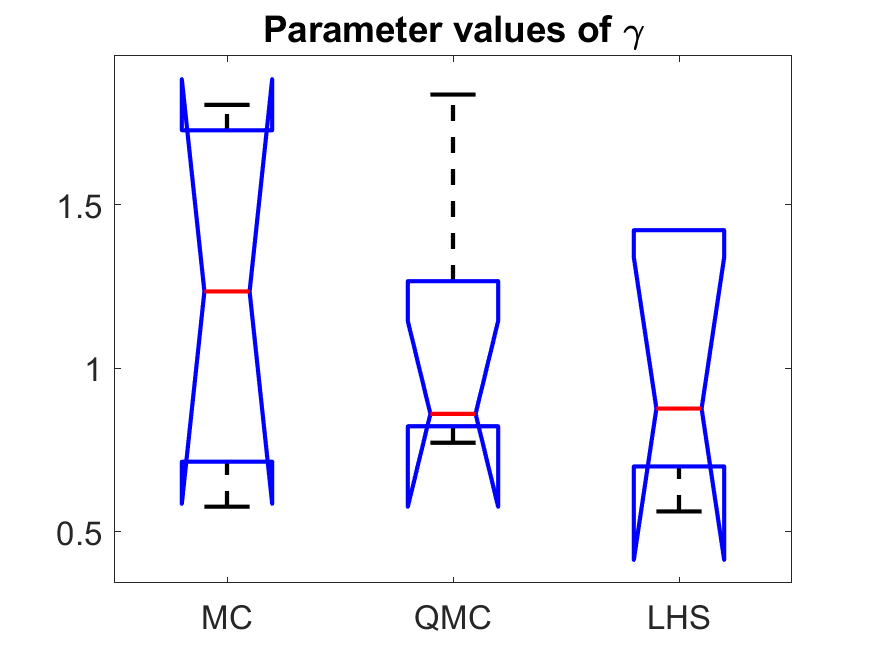}
\caption{{Boxplots of parameter values by MC, QMC, and LHS.}}
\label{Fig-boxplot-100}
\end{figure}

The parameter values of $\theta$, $\beta$ and $\gamma$ obtained by three different tuning methods can be visualized using the box plots, as shown in
Fig.~\ref{Fig-boxplot-100} where the red lines correspond the medians.

{For the values of $\theta$, the box plot does not show any significant differences in values because of the overlap of the value ranges at the 95\% confidence intervals. Similarly, the values of $\beta$ and $\gamma$ also indicated no significant differences. The box plots suggest that all parameters are within the allowed varying ranges. More comprehensive tests will be carried out to evaluate this further. In the rest of this paper, multiple tests using F-tests, Friedman tests and ANOVA will be carried out.}

\section{More Comprehensive Statistical Tests}

{Though the above t-tests can provide some insights about the mean objective values obtained by different tuning methods, these tests have focused on the mean values, not about their variances. To explore the possible variations of variances, F-tests will be used to test any potential differences in variances.
Non-parametric Friedman tests can also provide some insights from different perspectives.
For this reason, non-parametric Friedman tests will also be carried out to see if the same conclusions from t-tests are still valid. Furthermore, ANOVA can be considered as an extension of t-tests, and a full set of ANOVA tests are also carried out. These different statistical tests will look at the simulation data from different perspectives with different focuses and thus provide a fuller picture of the analysis of the results. }

\subsection{F-Test for Variances}

The t-tests are used for testing any possible significant differences in mean values, whereas Fisher's F-tests are for testing the potential differences in variances of two populations~\cite{schumacker2013ftest}. The main test statistic is the ratio of two sample variances
\be F=\frac{S_x^2}{S_y^2}, \ee
which obeys an F-distribution~\cite{mario2012elementary,schumacker2013ftest}.
All the p-values for the paired F-tests are summarized in Table~\ref{Tab-p-ft-100}.

\begin{table}[ht]
\begin{center}
\caption{{The p-values of paired F-tests.}}
\label{Tab-p-ft-100}
\begin{tabular}{|l|ccc|}
\hline
Function & MC vs QMC & MC vs LHS & QMC vs LHS \\
\hline
Sphere $f_1$ & 0.0798  &   0.6487  &   0.1854 \\
Robsenbrock $f_2$ & 0.0006  &   0.0052  &   0.4241 \\
Ackley $f_3$ & 0.8852 & $6.13 \times 10^{-13}$  & $9.57 \times 10^{-13}$ \\
Trid $f_4$ & 0.4547  &   0.3027  &   0.7724 \\
Spring $f_5$ & 0.3362  &   0.3293  &   0.0588 \\
Truss $f_6$ & 0.9942  & $2.47 \times 10^{-06}$ & $2.52 \times 10^{-06}$\\
\hline
\end{tabular}
\end{center}
\end{table}

The F-tests on the objectives values of the Sphere function, calculated using FA and the three tuning methods MC, QMC, and LHS show that the p-values for $f_1$ are greater than 0.05. The p-values are 0.0798, 0.6487, 0.1854 for MC versus QMC, MC versus LHS, and QMC versus LHS, respectively. These results indicate that fitness values calculated for the Sphere function using the three tuning methods display no major variation.

For $f_2$, the p-values are 0.0006, 0.0052, and 0.4241, respectively. Thus, the variations of the objective values are significantly different between MC and QMC, and MC and LHS, though there are not much differences between QMC and LHS. For $f_3$, the p-values are 0.8852, $6.13 \times 10^{-13}$, and $9.57 \times 10^{-13}$. This means that the variance of LHS is very different from MC and QMC, though there are no significant differences between MC and QMC.

{For $f_4$, the p-values are 0.4547, 0.3027, 0.7724. This indicates that the variations are not statistically significant in the fitness values obtained for the Trid function using the three tuning methods. For $f_5$, the observed p-values are 0.3362, 0.3293, 0.0588 from MC vs QMC, MC vs LHS and QMC vs LHS, respectively.
All the p-values are greater than 0.05, which indicates that the variations are not statistically significant in fitness values from different tuning methods for the spring design problem.

For $f_6$, the p-values obtained from pair-wise testing of MC vs QMC, MC vs LHS and QMC vs LHS are 0.9942, $2.47 \times 10^{-06}$, and $2.52 \times 10^{-06}$, respectively. This means that
no significant differences in variances are observed in the objective values obtained by MC and QMC for the truss sytem. However, the variances in the objective values obtained by LHS are statistically different from those by MC and QMC.

Based on the above hypothesis tests, functions with zero optimal objective values can have significant variations in objective values, even though their mean values are not significantly different. On the other hand, if the optimal objective values are not close to zero, the variations of the objective values obtained by different tuning methods are approximately the same order without any significant variations. This may indicate that the quality of the obtained solutions may depend on the type of optimization problem used for simulation. This may also suggest that the choice of the parameter tuning method should consider the type of problems to be solved.

In addition, many tests in the literature using Student's t-tests and other tests for function benchmarks with optima at zero may have potential problems because some conclusions in the literature may have over-estimated the differences in objective values, concerning the validation and testing of new metaheuristic algorithms.  Therefore, care should be taken when interpreting the results. In the rest of the paper, more statistical tests will be carried out. }

\subsection{Friedman Tests}

Friedman tests are non-parametric tests, which use ranking of data, rather than the exact values of the raw data~\cite{Zimmerman2010,Sedgwicktests2012,GeraldTests2021}.
Friedman tests are the extension of the sign tests, which evaluates possible effects  in mean values of columns or features. In addition, it is not required that data samples are normally distributed.

{For all the objective values of all benchmarks, non-parametric Friedman tests have been carried out.
The p-values from Friedman tests
are 0.8233, 0.6703, 0.2792, 0.4346, 0.6703, and 0.7275 for $f_1$, $f_2$, $f_3$, $f_4$, $f_5$, and  $f_6$, respectively.
Since all these p-values are greater than 0.05,
there are no significant differences among the tuning methods when using six different benchmarks.

For the tuned parameter values of $\theta$, $\beta$ and $\gamma$, the same types of Friedman tests
have been carried out. The p-value for $\theta$ values is 0.8465,
thus there are no significant differences among MC, QMC, and LHS. In addition,
for $\beta$ values, the p-value is 0.6065, which indicates no significant differences between all the three tuning methods. For $\gamma$ parameter values, the p-value obtained is 0.6065. This again means that  no significant differences
are observed among the three tuning methods. }

\subsection{ANOVA}

The analysis of variance (ANOVA) is an extension of the t-tests to compare the means of three or more samples, which can also take variances into consideration~\cite{Zimmerman2010,Quirk2012,Wilcox2022}.

{For the objective values, the standard two-way ANOVA has been carried out, and both row effect and column effect are considered. From these tests, the smallest p-values are 0.7390, 0.2548, 0.3136, 0.5637, 0.6493, and 0.6475 for $f_1$, $f_2$, $f_3$, $f_4$, $f_5$, and $f_6$, respectively. All these p-values are greater than 0.05, which indicates that there are no significant differences among the solution quality of different benchmarks.

For the ANOVA analysis of $\theta$ values, the p-value for $\theta$ is 0.8936, which may indicate that no significant differences exist in the values of $\theta$ obtained by different tuning methods.
Similarly, for $\beta$ values, the p-valve is 0.3605, which again shows no statistically significant differences in parameter values among three different tuning methods. Finally, for $\gamma$ values, the p-value is 0.7157, which also suggests that there are no significant differences.

Based on the above observations and hypotheses tests, it can be concluded that the tuned values of the three parameters in the FA are largely independent of the tuning methods used. Thus, the FA can be tuned by any of the three tuning methods and the performance of the tuned FA can be equally good. This also implies that the FA is flexible and its parameter values are insensitive to the tuning methods, which is a noticeable advantage of the algorithm for solving optimization problems.
}

\section{Concluding Remarks and Further Research}

{Different methods for tuning parameters in nature-inspired algorithms have been investigated.
The three parameters of the FA have been tuned by using three different methods for parameter tuning, including the MC, QMC and LHS. Six benchmarks with diverse properties of convexity and constraints have been used to test the sensitivity of the parameter values and their possible influence on performance of the FA and possible difference arising from different tuning methods. Numerical experiments, followed by comprehensive statistical tests, show that there are no significant differences in both solution quality in terms of objective values and the tuned parameter values of the FA. }

The mean values of the objective functions are not significant, whatever the tuning methods are used for tuning the FA parameters, which has been confirmed by the t-tests. The variances have been also been analyzed and tested using F-tests and ANOVA. Again no significant differences have been found. In addition, the non-parametric tests by the Friedman test
also confirm such conclusions, even though the underlying parameter values and their variations may not obey normal distributions.  These tests indicate that the FA can perform equally well, independent of which of the three parameter tuning methods used.

One possible interesting issue that was observed in the above analysis is that different types of benchmarks, especially the location of optima, may have some minor effect on the variance of the objective values, though the solution quality and their means remain essentially the same. This point can be further explored in the future research.

{As a possible extension, this current work can be extended to tune other algorithms with more parameters, such as the unified generalized evolutionary metaheuristic (GEM) algorithm with seven parameters~\cite{Yang2024} and many other algorithms~\cite{Rajwar2023}. In addition, the Bayesian optimization can be used for hyper-paremeter tuning and  optimization~\cite{Garnett2023bayesian}. Furthermore, some theoretical analysis of the population variance of the FA can be carried out in the similar way as those for differential evolution~\cite{Zaharie2009}, which may gain some key insights into the parameter settings. All these topics will be explored further in future studies.

\bibliographystyle{plain}

\end{document}